\documentclass[aps,twocolumn,prl]{revtex4}
\usepackage{graphicx}
\usepackage{amsmath}
\begin{document}
\date{\today}
\title{Mechanical Stability of a Strongly-Interacting Fermi Gas of Atoms}
\author{M. E. Gehm}
\author{S. L. Hemmer}
\author{S. R. Granade}
\author{K. M. O'Hara}
\author{J. E. Thomas}
 \affiliation{Physics Department, Duke University, Durham, North Carolina
27708-0305} \pacs{03.75.Ss, 32.80.Pj}

\begin{abstract}
A strongly-attractive, two-component Fermi gas of atoms exhibits
universal behavior and should be mechanically stable as a
consequence of the quantum mechanical requirement of unitarity.
This requirement limits the maximum attractive force to a value
smaller than that of the outward Fermi pressure. To experimentally
demonstrate this stability, we use all-optical methods to produce
a highly degenerate, two-component gas of $^6$Li atoms in an
applied magnetic field near a Feshbach resonance, where  strong
interactions are observed. We find that the gas is stable at
densities far exceeding that predicted previously for the onset of
mechanical instability. Further, we provide a
temperature-corrected measurement of an important, universal,
many-body parameter  which determines the stability---the mean
field contribution to the chemical potential in units of the local
Fermi energy.
 \end{abstract}
 \maketitle

Strongly interacting Fermi systems are expected to exhibit
universal behavior~\cite{Heiselberg}. In atomic gases, such strong
forces can be produced in the vicinity of a Feshbach resonance,
where a bound molecular state in a closed exit channel is
magnetically tuned into coincidence with the total energy of a
pair of colliding particles~\cite{Verhaar}. In this case, the zero
energy scattering length $a_S$, which characterizes the
interactions at low temperature, can be tuned through $\pm\infty$.
For very large values of $|a_S|$, the important properties of the
system (e.g. the effective mean field potential, the collision
rate, the superfluid transition temperature, etc.) are predicted
to lose their dependence on the magnitude and sign of $a_S$, and
instead become proportional to the Fermi energy with different
universal proportionality constants. For this reason, tabletop
experiments with strongly interacting atomic Fermi gases can
provide measurements that are relevant to all strongly interacting
Fermi systems~\cite{Heiselberg,Steele}, thus impacting theories in
intellectual disciplines outside atomic physics, including
materials science and condensed matter physics
(superconductivity), nuclear physics (nuclear matter), high-energy
physics (effective theories of the strong interactions), and
astrophysics (compact stellar objects).

In a gas of degenerate fermions, compression of the gas is
resisted by the Pauli exclusion principle, leading to an effective
pressure known as the Fermi pressure or degeneracy pressure. The
Fermi pressure plays an important role in nature, providing, for
example, the outward force which stabilizes neutron stars against
gravitational collapse. Because of the Fermi pressure, a confined
cloud of degenerate fermions is always larger than a cloud of
bosons with an equivalent temperature and particle number. This
effect has been directly observed in two elegant
experiments~\cite{Truscott,Salomon}.

An important question is: At what point, if at all, do strong
attractive interactions overcome the Fermi pressure? Previously,
it was predicted that an atomic Fermi gas could  become
mechanically unstable for sufficiently large attractive
interactions~\cite{Houbiers}. If true, this prediction presents a
possible roadblock to attempts to use a Feshbach resonance to
produce a high-temperature
superfluid~\cite{Holland,Timmermans,Kokkelmans}, as the gas should
be unstable in the required regime. Recently, however, it has
become apparent that the gas may indeed be stable. Heiselberg has
estimated the maximum ratio of the attractive potential  to the
local Fermi energy for a strongly interacting Fermi gas. Using a
self-consistent many-body approach, he finds that a two-component
Fermi gas is mechanically stable, although a gas with more than
two components is not~\cite{Heiselberg}. A similar conclusion can
also be drawn from an examination of the compressibility of a
strongly attractive two-component gas~\cite{Kokkelmans}.

In this Letter, we first show heuristically that the maximum
inward force arising from attractive interactions in a two-state
gas of atomic fermions  is limited by quantum mechanics to a value
less than the outward Fermi pressure. We then demonstrate this
idea experimentally by directly cooling a two-component Fermi gas
of $^6$Li atoms to high degeneracy in an optical trap at magnetic
fields near a Feshbach resonance, where strong interactions are
observed~\cite{OharaScience}. The highest densities obtained in
the experiments far exceed those predicted
previously~\cite{Houbiers} for the onset of mechanical
instability. Imaging the cloud both in the trap and after abrupt
release, we find no evidence of instability. To quantitatively
describe the stability, we present a measurement of the relevant
universal many-body parameter $\beta$---the mean-field
contribution to the chemical potential in units of the local Fermi
energy. In Ref.~\cite{OharaScience}, we provided a first estimate
of $\beta$ assuming a zero temperature gas. Here, we present a
method to properly account for the non-zero temperature of the
gas. This revised  measurement of $\beta$  is in good agreement
with a prediction of~\cite{Heiselberg}.

The mechanical stability of the gas can be understood through a
heuristic discussion of the forces acting on a gas of fermions in
a 50-50 mixture of two spin components. The equation of state for
a normal, zero-temperature Fermi gas is~\cite{Stringari}
\begin{equation}
\epsilon_F(\mathbf{x})+
U_{\text{MF}}(\mathbf{x})+U_{\text{trap}}(\mathbf{x})=\mu ,
\label{eq:chempot}
\end{equation}
where $\mu$ is the chemical potential, $\epsilon_F(\mathbf{x})$ is
the local Fermi energy, $U_{\text{MF}}(\mathbf{x})$ is the mean
field contribution to the chemical potential, and
$U_{\text{trap}}(\mathbf{x})$ is the trap potential. In the local
density approximation,
$\epsilon_F(\mathbf{x})=\hbar^2\,k_F^2(\mathbf{x})/(2M)$, where
$k_F$ is the local Fermi wavevector, which is related to the
density according to $n(\mathbf{x})=k_F^3(\mathbf{
x})/(6\pi^2)$~\cite{Butts}.

For each spin component, there is an effective outward
force~\cite{fermiforce} arising from the Fermi pressure
$P_{\text{Fermi}}=2n(\mathbf{x})\,\epsilon_F(\mathbf{x})/5$, even
in the absence of interatomic interactions.  The outward force per
unit volume is $-\mathbf{\nabla}P_{\text{Fermi}}$. The
corresponding force per particle is $-(\mathbf{ \nabla}P_{\text{
Fermi}})/n$. The local force per particle is then  easily shown to
be
\begin{equation} \mathbf{F}_{\text{Fermi}}=-{\boldsymbol
\nabla}\epsilon_F(\mathbf{x}), \label{eq:fermiforce}
\end{equation}
 Since the density decreases with
distance from the center, $\mathbf{ F}_{\text{Fermi}}$ is outward.

Two body scattering interactions make a contribution
$U_{\text{MF}}$ to the chemical potential, which is given in the
mean field approximation by,
\begin{equation}
U_{\text{MF}}(\mathbf{x})=\frac{4\pi\hbar^2\,a_{\text{eff}}}{M}\,n(\mathbf{
x}), \label{eq:meanfield1}
\end{equation}
where $a_{\text{eff}}$ is an effective scattering length, which
generally depends on  the local thermal average of a
momentum-dependent scattering amplitude. The potential is
attractive when $a_{\text{eff}}<0$, and repulsive when
$a_{\text{eff}}>0$. The local force per particle arising from the
mean field contribution is just
\begin{equation}
\mathbf{F}_{\text{MF}}=-{\boldsymbol\nabla}U_{\text{MF}}(\mathbf{x})
 . \label{eq:meanforce}
\end{equation}

If we assume that $a_{\text{eff}}$ is energy-independent and equal
to $a_S$ for $a_S<0$, it is easy to show that the gas becomes
unstable for suitably large values of $a_S$. In this case, the
inward force from the mean-field potential,
$\mathbf{F}_{\text{MF}}\propto \boldsymbol\nabla n(\mathbf{x})$
while the outward force from the Fermi pressure,
$\mathbf{F}_{\text{Fermi}}\propto \boldsymbol\nabla
n^{2/3}(\mathbf{x})$. Then the inward force exceeds the outward
when $|\mathbf{F}_{\text{MF}}|>|\mathbf{F}_{\text{Fermi}}|$, i.e.,
when $k_F\,|a_{\text{eff}}|>\pi /2$. The corresponding density for
mechanical instability satisfies $n>n_0$ where $n_0
=\pi/(48\,|a_S|^3)$. This result matches the previous prediction
of~\cite{Houbiers} which was derived via a more rigorous
calculation.

The assumption that $a_{\text{eff}}$ is energy-independent,
however, is only valid if $k_F\,|a_S|\ll 1$. Outside that regime,
this assumption violates the quantum mechanical requirement of
unitarity. At intermediate densities~\cite{Heiselberg}, where
$a_S\gg k_F^{-1}\propto n^{-1/3}\gg R$, with $R$ the range of the
collision potential, two-body scattering is dominant, but the mean
field interaction is proportional to a momentum-dependent two-body
T-matrix element, which in turn is proportional to the scattering
amplitude $f(k)$. It is well known that $f$ has a magnitude which
is limited by unitarity to a maximum of $1/k$~\cite{Landau}.
Hence, one expects that in a zero-temperature Fermi gas, one
should use an effective scattering length in
Eq.~\ref{eq:meanfield1} with a maximum magnitude on the order of
$a_{\text{eff}}=1/k_F$. For $a_{\text{eff}} = -1/k_F$, it is easy
to show that the mean field contribution to the chemical potential
is~\cite{OharaScience},
\begin{equation}
U_{\text{MF}}(\mathbf{ x}) =\beta\epsilon_F(\mathbf{ x}),
\label{eq:meanfield}
\end{equation}
where the universal parameter $\beta$ has a maximum value of $-
4/(3\pi)=-0.42$ in this heuristic treatment. In this limit, both
the inward force, $\mathbf{F}_{\text{MF}}$ and the outward force,
$\mathbf{F}_{\text{Fermi}}$ are proportional to the same power of
the atomic density. The ratio of their magnitudes, however, is
given by
$|\mathbf{F}_{\text{MF}}|/|\mathbf{F}_{\text{Fermi}}|=|\beta |<1$.
Thus, one finds that a two-component Fermi gas is mechanically
stable as a result of the quantum mechanical requirement of
unitarity. This is in  contrast to attractive Bose
gases~\cite{Hulet,Wieman} and Bose-Fermi mixtures~\cite{Inguscio}
which exhibit dramatic instabilities.

A possible criticism of our heuristic argument is that the
two-body relative wavevector $k$ lies in the range  $0\leq k\leq
k_F$. Since $f$ increases as $k$ decreases, $f(k)$ should be
averaged over $W(\mathbf{ k})$, the probability distribution for
$\mathbf{k}$, where $\int d\mathbf{k}\,W(\mathbf{k})=1$. For a
noninteracting gas at zero temperature, one can show
\begin{equation}
W(\mathbf{ k})=\frac{6}{\pi\,
k_F^3}\left[1-\frac{3}{2}\frac{k}{k_F}+\frac{1}{2}\left(\frac{k}{k_F}\right)^3\right]\Theta
(k_F-k) , \label{eq:relkdistrib}
\end{equation}
where $k=|\mathbf{ k}|$.
 Near a Feshbach resonance, where $|a_S|\gg R$, we assume that $f(k)$ takes the
form expected for a zero energy resonance, and that the effective
scattering length is determined by the real part of $-f$, i.e.,
$a_{\text{eff}}=\langle a_S/(1+k^2\,a_S^2)\rangle$, where $\langle
...\rangle$ denotes averaging with Eq.~\ref{eq:relkdistrib}. We
find that $|a_{\text{eff}}|$ has a maximum value of $1.05/k_F$
when $k_F\,|a_S|=2$, and decreases slowly for $k_F|a_S|>1$,
reaching $0.6/k_F$ at $k_F|a_S|=10$.

Our heuristic estimate, which neglects the effects of the
interactions on the free particle wavefunctions,  yields a value
of $\beta$ comparable to that estimated  by
Heiselberg~\cite{Heiselberg}. In contrast to our heuristic
approach, the self consistent many-body approach of
Ref.~\cite{Heiselberg} also predicts that $\beta$  is
 independent of the sign and magnitude of
$k_F\,a_S$ in the intermediate density limit where $k_F\,|a_S|\gg
1 $. Hence, for strongly interacting fermions, $\beta$ is always
negative and  the effective mean field interaction should always
be attractive.

To demonstrate that a strongly attractive, two spin-component
Fermi gas of atoms is stable,  we employ a 50-50 mixture of the
two lowest hyperfine states of $^6$Li, i.e., the $|F=1/2,M=\pm
1/2\rangle$ states in the low field basis. This mixture has a
predicted broad Feshbach resonance at 860 G~\cite{Elastic,
Zerocross}. A magnetic field of 910 G is applied to produce a very
large and negative zero-energy scattering length.

 The gas mixture is prepared and  rapidly cooled to
degeneracy at 910 G by forced evaporation in our ultrastable
CO$_2$ laser trap, as described for our previous
experiments~\cite{OharaScience}. For our trap,
$\omega_\perp=2\pi\times (6625\pm 50\,\text{Hz} )$,
$\omega_z=2\pi\times (230\pm 20\,\text{Hz})$, and
$\bar{\omega}\equiv (\omega_\perp^2\omega_z)^{1/3}=2\pi\times
(2160\pm 65\, \text{Hz})$. Following evaporation, the trap is
recompressed to full depth over 0.5 s, and allowed to remain for
an additional 0.5 s, to ensure thermal equilibrium. The CO$_2$
laser power is then extinguished and the gas is imaged as
described in our previous paper~\cite{OharaScience}. After
release, the gas expands hydrodynamically, rapidly increasing in
the transverse dimension while remaining nearly stationary in the
axial direction~\cite{OharaScience}.

We determine the  number of atoms by numerically integrating the
column density (Table~\ref{table:1}). For a typical number,
$N\simeq 8.0\times 10^4$ atoms per state, the corresponding Fermi
density for the experiments is calculated to be $n_F\simeq
4.8\times 10^{13}/{\text{cm}}^{3}$ per state~\cite{Butts}. The
density $n_F$, can be compared to the density
$n_0=\pi/(48\,|a_S|^3)$ predicted for the onset of mechanical
instability, assuming a momentum-independent scattering length.
Using the best available molecular potentials, which are
constrained by measurements of the zero crossing in the s-wave
scattering length~\cite{Zerocross,Grimm}, the zero energy
scattering length $a_S$ is estimated to be $\simeq -10^4\,a_0$ at
910 G ($a_0=0.53\times 10^{-8}$ cm), within a factor of
two~\cite{Vanessa}. Then $n_0=4.4\times 10^{11}/{\text{cm}}^3$,
showing that the density $n_F$  exceeds $n_0$ by one to two orders
of magnitude.

Although the molecular potentials are constrained by the measured
zero crossing in the scattering length~\cite{Zerocross,Grimm}, the
precise location of the  Feshbach resonance still may be
incorrect. Since the scattering length has not been directly
measured, it is possible that the attractive potential is not as
large as expected. However, a simple argument based on our
previous experiments shows that $a_S$ must be
large~\cite{OharaScience}. We observe similar hydrodynamic
expansion  after release  from either full trap depth $U_0$ or
from a reduced trap depth of $U_0/100$. In the latter case, we
estimate that $k_F^{-1}=4000\,a_0$. Since the observed
hydrodynamic expansion appears to be independent of the trap
depth, as expected for a unitarity limited interaction, we
conclude that in the shallow trap we must have $|k_F\,a_S|>1$.
Then we must have $|a_S|>4000\,a_0$ and $n_0\leq 6.9\times
10^{12}/{\text{cm}}^3$.

Thus, we have clearly demonstrated that the gas is mechanically
stable in the intermediate density regime, where $|a_S|$ is large
compared to the interparticle spacing. The remainder of this paper
provides a revised estimate of the universal parameter $\beta$
which quantitatively determines the stability of the gas. In our
previous estimate~\cite{OharaScience}, we assumed a zero
temperature gas and calculated $\beta$ from an estimate of the
release energy. There, we obtained a value of $\beta$
substantially smaller in magnitude than
predicted~\cite{Heiselberg}. Here, we determine $\beta$ from the
transverse spatial widths of the expanding gas and properly
include both finite temperature and hydrodynamic scaling effects.

\begin{figure}
\begin{center}\
\includegraphics[height=1.75in,width=2.5in]{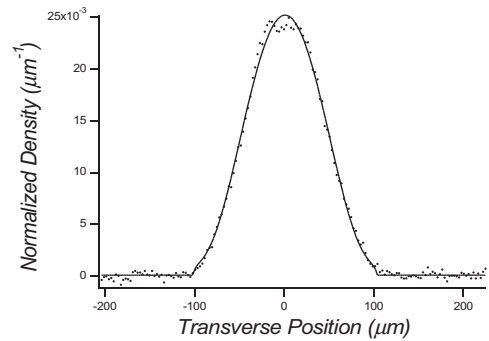}
\end{center}
\caption{\label{fig:2}Transverse spatial profile of the expanding
cloud at 600 $\mu$s after release.}
\end{figure}

The one dimensional  transverse spatial profiles after expansion
for 0.4-0.8 ms are very well fit by normalized finite-temperature
Thomas-Fermi (T-F) distributions
 which determine the width $\sigma_x$ as well as the ratio of the temperature
to the Fermi temperature, $T/T_F$.  We expect that $T/T_F$ is
approximately constant during the expansion, since we observe
hydrodynamic scaling of the transverse radii consistent with an
effective potential $\propto n^{2/3}$~\cite{OharaScience},
suggesting an adiabatic process. The T-F  shape is not
unreasonable despite a potentially large mean field interaction,
since  the mean field contribution to the chemical potential,
Eq.~\ref{eq:meanfield} is proportional to the local Fermi energy.
In this case, assuming Eq.~\ref{eq:chempot}  for a normal
degenerate gas at zero temperature~\cite{Stringari}, it is easy to
show that the mean field should simply scale the Fermi energy of
the trapped cloud without changing the shape from that of a T-F
distribution~\cite{OharaScience}. The zero temperature spatial
distribution then corresponds to that of a harmonic oscillator
potential, with the frequencies scaled so that
$\omega_i'=\omega_i/\sqrt{1+\beta}$, $i=x,y,z$. Hydrodynamic
expansion then preserves the shape of the trapped atom spatial
distribution~\cite{OharaScience,Stringari}. Assuming that $T/T_F$
is small, one expects that a finite temperature spatial
distribution for a harmonic potential is a reasonable
approximation. A Sommerfeld expansion~\cite{AshcroftMermin} of the
normalized density for $|x|\leq\sigma_x$ yields
\begin{equation}
n(x)/N=\frac{16}{5\pi\sigma_x}\left[f_0(x)+5\pi^2\,(T/T_F)^2\,f_2(x)\right],
\label{eq:density}
\end{equation}
where $f_0(x)=g^5(x)$, $f_2(x)=g(x)/8-g^3(x)/6$, and
$g(x)=\sqrt{1-x^2/\sigma_x^2}$.

Eq.~\ref{eq:density} is used to fit the measured transverse
spatial distributions, Fig.~\ref{fig:2}. We obtain the results
given in Table~\ref{table:1} for four trials each at expansion
times $t$ of 0.6 ms and 0.8 ms. Note that Eq.~\ref{eq:density}
begins to break down near $|x|\simeq \sigma_x$ for $T/T_F>0.15$.
However, an exact treatment using polylogarithm functions yields
similar results even for our highest temperatures where
$T/T_F\simeq 0.18$.

\begin{table}[!]
\caption{\label{table:1}Value of $\beta$: $t$ is the expansion
time, $N$ is the number of atoms per state,  $\sigma_{xF}$ is
Fermi radius without the mean field interaction, $\sigma_x(t)$ is
the measured Fermi radius obtained from the finite temperature fit
to the data using Eq.~\ref{eq:density}. }
\begin{ruledtabular}
\begin{tabular}{|c|c|c|c|c|c|}

$t$($\mu$s)&$N$&$\sigma_{xF}$ ($\mu$m)&$T/T_F$&$\sigma_x(t)$ ($\mu$m)&$\beta$\\
\hline
600&66,000&3.49&0.128&99&-0.237\\
&86,400&3.65&0.144&102&-0.220\\
&84,300&3.63&0.140&101&-0.234\\
&80,200&3.60&0.141&101&-0.208\\
800&67,200&3.50&0.146&133&-0.176\\
&87,000&3.65&0.179&131&-0.344\\
&70,400&3.53&0.151&130&-0.273\\
&82,000&3.62&0.183&128&-0.382\\
\end{tabular}
\end{ruledtabular}
\end{table}

The value of the parameter $\beta$ can be determined from the
transverse radii  of the trapped cloud,
$\sigma_x(0)=\sqrt{2\epsilon_F'/M\omega^{'2}_{\perp}}$, where
$\epsilon_F'=\hbar\bar{\omega}'(6N)^{1/3}$ is the Fermi energy
including the mean field contribution. Then,
$\sigma_x(0)=(1+\beta)^{1/4}\sigma_{xF}$~\cite{Stringari3}, where
$\sigma_{xF}=\sqrt{2\epsilon_F/(M\omega_\perp^2)}$, and
$\epsilon_F=\hbar\bar{\omega}\,(6N)^{1/3}$ is the Fermi energy in
the absence of interactions.  As we have pointed out previously,
the observed anisotropic expansion can arise from
unitarity-limited collisional hydrodynamics or superfluid
hydrodynamics~\cite{OharaScience}. In either case, the effective
potential is $\propto n^{2/3}$, and we can assume
$\sigma_x(t)=\sigma_x(0)\,b_x(t)$~\cite{OharaScience,Stringari}.
Hence,
\begin{equation}
\beta=\left(\frac{\sigma_x(t)}{b_x(t)\sigma_{xF}}\right)^4-1,
\label{eq:beta}
\end{equation}
For our trap parameters, $b_x(0.6\,\text{ms})=29.74$ and
$b_x(0.8\,\text{ms})=39.88$. The scale factor $b_x(t)$ properly
includes the spatial anisotropy of the  expansion and the correct
hydrodynamic scaling. We obtain from Table~\ref{table:1} an
average value $\beta=-0.26\pm 0.07$.

This result is consistent with that obtained from the measured
transverse release energy. The release energy can be initially
calculated from the zero temperature T-F fits~\cite{OharaScience}
or calculated numerically from second moment of the density. Since
$T/T_F\simeq 0.15$ from Table~\ref{table:1}, the measured release
energy is larger than the zero temperature release energy by a
factor $\eta\simeq 1+(2\pi^2/3)(T/T_F)^2=1.15$. Reducing the
release energy by a factor $\eta$ and using the method of
Ref.~\cite{OharaScience}, yields an average value of $\beta$
consistent with that given above.

The measured average value of $\beta=-0.26\pm 0.07$ can be
compared with predictions of the energy per particle for the
conditions of our experiment, where $k_Fa_S=-7.4$. For our
equation of state, the sum of the local kinetic and mean field
energy per particle is
$(3/5)(1+\beta)\epsilon_F(\mathbf{x})$~\cite{Stringari2}. In
Ref.~\cite{Heiselberg}, Heiselberg provides two estimates for the
total energy per particle. His Eq. 11, based on the Galitski
equations, yields $\beta_{\text{calc}}=-0.54$~\cite{Error}. Using
the Wigner-Seitz cell approximation,  his Eq. 14  provides an
alternative estimate, $\beta_{\text{calc}}=-0.33$. An independent
calculation by Steele~\cite{Steele} using effective field theory
yields $\beta =-0.46$. The theoretical predictions are for zero
temperature, and the second is in reasonable agreement with our
measurements including the temperature correction to the spatial
distributions. However, the predictions do not include the
additional temperature dependence of $\beta$ which arises from a
thermal average. Since this important universal many-body
parameter can now be experimentally measured, further refinement
of both the experimental measurements and the theory is worthwhile
and may permit  the observation of the superfluid corrections to
$\beta$. In addition, mixtures of the three lowest hyperfine
states of $^6$Li may permit observation of predicted mechanical
instabilities in a three-component Fermi gas~\cite{Heiselberg}.

This research is supported by the Physics Division of the Army
Research Office  and the Fundamental Physics in Microgravity
Research program of the National Aeronautics and Space
Administration, the  Chemical Sciences, Geosciences and
Biosciences Division of the Office of Basic Energy Sciences,
Office of Science, U. S. Department of Energy and by the National
Science Foundation.



\begin{thebibliography}{99}
\bibitem{fermiforce}
Note that Fermi energy in the chemical potential of a
noninteracting trapped gas produces an effective outward force
which balances the inward trap force to determine the local
density. However, a noninteracting gas expands ballistically after
release from the trap.
\bibitem{Stringari2}Note that the equation of state is obtained
from the total energy by functional differentiation with respect
to the density, yielding the minimum energy state. S. Stringari,
private communication.
\end{thebibliography}

\end{document}